\documentclass[12pt,aip,jcp,unsortedaddress]{revtex4-1}
\usepackage[utf8]{inputenc}
\usepackage{amsmath}
\usepackage{graphicx}
\usepackage{url}

\usepackage{color}
\usepackage[nolist]{acronym}
\begin{acronym}
 \acro{qd}[QD]{quantum dynamics}
\end{acronym}

\DeclareMathOperator\erf{erf}
\begin{document}
\title{Revisiting an approximation in the Wilson G-matrix formalism and its impact on molecular quantum dynamics}
\author{Julius P. P. Zauleck}
\affiliation{Department Chemie, Ludwig-Maximilians-Universit\"{a}t M\"{u}nchen, Butenandtstr. 13, 81377 M\"{u}nchen, Germany}
\author{Sebastian Thallmair}
\affiliation{Groningen Biomolecular Sciences and Biotechnology Institute and The Zernike Institute for Advanced Material, University of Groningen, Nijenborgh 7, 9747 AG Groningen, Netherlands}
\author{Regina de Vivie-Riedle}
\affiliation{Department Chemie, Ludwig-Maximilians-Universit\"{a}t M\"{u}nchen, Butenandtstr. 13, 81377 M\"{u}nchen, Germany}

\begin{abstract}
Quantum dynamics simulations of reactive molecular processes are commonly performed in a low-dimensional space spanned by highly optimized reactive coordinates. Usually, these sets of reactive coordinates consist of non-linear coordinates. The Wilson G-matrix formalism allows to formulate the Hamiltonian in arbitrary coordinates. In our present work, we revisit an approximation in this formalism, namely the assumption that the Jacobian determinant is constant. We show that the approximation can introduce an error and illustrate it for a harmonic oscillator. Finally, we present a strategy to prevent this error.
\end{abstract}

\maketitle

The reduction of coordinates is an important aspect in molecular \ac{qd} \cite{Nyman2013, Nyman2014}. This naturally comes along with the search for the most important coordinates to describe the molecular process of interest \cite{Yagi2001, Iung2006, Joubert2012, Thallmair2016a, Zauleck2016, Li2017, Zauleck2017}. Often Z-matrix coordinates or linear combinations of them are chosen because they represent molecular motions in an intuitive way. As these Z-matrix coordinates include e.g. angles they are non-linear which in turn complicates the kinetic part of the Hamiltonian. \cite{Meyer2007} A straightforward way to deal with this is the Wilson G-matrix formalism \cite{Podolsky1928, Wilson1955, Schaad1989}. It allows to transform the kinetic Hamiltonian from the Cartesian space to arbitrary coordinates and offers a simple way to reduce the degrees of freedom using a block matrix inversion. The G-matrix gives access to the kinetic couplings along the non-linear coordinates and its diagonal elements can be interpreted as position dependent inverse reduced mass \cite{Thallmair2016a}. 

The G-matrix formalism was first introduced by B. Podolsky \cite{Podolsky1928} to reformulate the quantum mechanical Hamiltonian. Later, it became a common tool for the description and interpretation of vibrational spectra where the Hamiltonian had to be expressed in normal mode coordinates \cite{Wilson1939, Wilson1955, Ribeiro2005}. In its common formulation, the G-matrix formalism comes along with an approximation assuming that the Jacobian determinant $j=\textrm{det}\left|\mathbf{J}\right|$ is constant. This approximation is usually justified by arguing that the position dependence of the G-matrix is more pronounced than that of the Jacobian determinant. 

To the best of our knowledge, this approximation and its effect has never been checked for molecular \ac{qd}.
Thus, we revisit this approximation in our present work. While a transformation into linear coordinates, e.g. normal modes that are often used for small displacements around equilibrium structures, assures a constant Jacobian determinant $j$ and introduces no error, this is not the case for non-linear coordinates. We investigate if a certain setup of the chosen subspace coordinates can reduce the error due to approximating the Jacobian determinant as constant or if it might be even possible to avoid this error at all. 
In doing so, we first introduce the approximation typically applied within the G-matrix formalism. Then, we demonstrate for an example of a simple coordinate transformation for a harmonic oscillator when the approximation leads to an obvious error. Finally, we present a strategy how to deal with a molecular system for which the dimensionality reduction and the approximation of a constant Jacobian determinant are applied at the same time. Note that all equations are written in atomic units.

While there exist alternative approaches that ensure a correct treatment of the Jacobian determinant, they rely on the knowledge of the complete transformation between the Cartesian coordinates and a full set of reactive and non-reactive coordinates \cite{Nauts1987,Gatti1997}. This is in contrast to the G-matrix formalism, where only the transformation between the small set of reactive coordinates and the Cartesian coordinates is required and thus clearly has practical advantages. The problem of the correct treatment is also not unique to the approximation that is the focus of this work. It appears whenever a Jacobian determinant is part of a reduced dimensional description, which is likely always the case for the transformation of the kinetic energy operator. Our conclusions also apply to these cases.

To describe the time evolution of the nuclear wavefunction $\Psi(x,t)$ of a molecular system, the time-dependent Schr\"{o}dinger equation of the nuclei has to be solved:
\begin{align}
 i \frac{\partial}{\partial t} \Psi\left(x,t\right) &= \hat H \Psi\left(x,t\right) \label{eq:sg} \\
 &= \left(\hat T + \hat V\right) \Psi\left(x,t\right)  \text{ .}
\end{align}
with the Hamiltonian $\hat H$ being the sum of the kinetic operator $\hat T$ and the potential operator $\hat V$. For an appropriate description of molecular processes, the Schr\"{o}dinger equation is commonly transformed into reactive coordinates $\mathbf q$ \cite{Schaad1989, Tannor2007}. The multiplicative potential operator $\hat V$ can be transformed easily. This is not the case for the kinetic operator $\hat T$. The Wilson G-matrix formalism opens a way to reformulate the kinetic energy operator in a general set of coordinates. Its exact formulation is given by
\begin{align}
\hat T_\mathbf{q} &= -\frac{1}{2} \sum_{r=1}^{N_{\mathbf{q}}} \sum_{s=1}^{N_{\mathbf{q}}}j^{-\frac{1}{2}} \frac{\partial}{\partial q_r}\left[ j\mathbf{G}_{rs} \frac{\partial}{\partial q_s} j^{-\frac{1}{2}}\right] 
\label{eq1}
\end{align}
with the G-matrix elements $\mathbf{G}_{rs}$ being
\begin{align}
 \mathbf{G}_{rs} = \sum_{i=1}^{3N} \frac{1}{m_i} \frac{\partial q_r}{\partial x_i} \frac{\partial q_s}{\partial x_i} \text{ .}
\end{align}
Assuming that the Jacobian determinant $j$ is constant, all derivatives acting on $j$ can be ignored \cite{Alexandrov1998}, resulting in
\begin{align}
\hat T_\mathbf{q} &\approx -\frac{1}{2} \sum_{r=1}^{N_{\mathbf{q}}} \sum_{s=1}^{N_{\mathbf{q}}} \frac{\partial}{\partial q_r}\left[ \mathbf{G}_{rs} \frac{\partial}{\partial q_s} \right] \text{ .}
\label{eq2}
\end{align}

This approximation is often disregarded and thus not treated carefully enough which may result in a misunderstanding in grid based approaches for \ac{qd}. It might lead to sets of reactive coordinates $\mathbf{q}$ which inherit an error that can be easily removed.

\begin{figure} [ht!]
\begin{center}
   \includegraphics[width=0.6\columnwidth,keepaspectratio=true]{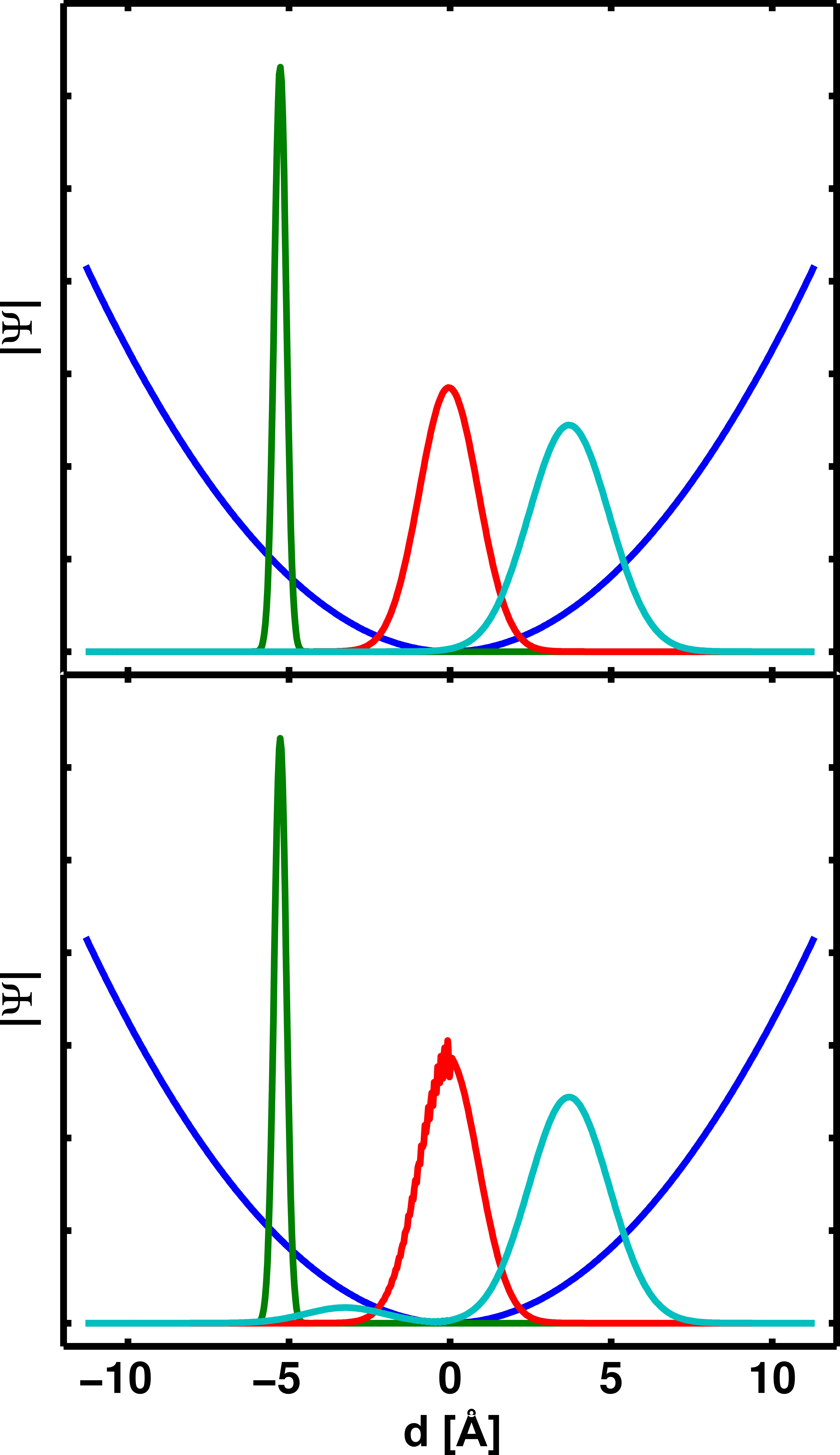} 
\end{center}
  \caption{Error introduced by approximating a constant Jacobian determinant $j$. The time evolution of a wavepacket inside a harmonic potential (blue) is shown at time 0 fs (green), 30 fs (red) and 54 fs (cyan). The upper panel shows the correct propagation on a Cartesian coordinate $x$, whereas the lower panel shows a erroneous propagation according to Eq. \ref{eq2} on a nonlinear coordinate $y$. For visualization consistent with the coordinate $x$, the propagation on $y$ in the lower panel is illustrated according to the transformation $\Psi(y)j(y)^{-\frac{1}{2}}$.}
  \label{fig:error}
\end{figure}

Figure \ref{fig:error} illustrates a simple scenario where this approximation causes a notable error. Here, a Gaussian wavepacket of the mass of a proton with a small initial momentum is propagated inside a harmonic potential (blue) using the Split-Operator method \cite{Feit1982}. The upper panel of Figure \ref{fig:error} shows the correct time evolution of the system in a Cartesian coordinate $x$ (or in other coordinates according to Eq. \ref{eq1}). The lower panel shows the time evolution on a coordinate $y$ with non-constant Jacobian determinant $j$ with respect to $x$ using the kinetic energy operator of Eq. \ref{eq2}. To introduce a significant nonlinearity around 0 {\AA}, the coordinate transformation is given by $x=a\erf(by)+y$ with the error function $\erf$, $a=-0.4$ and $b=1.5$. This leads to a compressed coordinate $y$ with respect to $x$ around 0 {\AA}. The propagation is shown at time 0 fs (green), 30 fs (red) and 54 fs (cyan). It can be seen that the error is introduced as the wavepacket crosses the nonlinearity in $y$ and the resulting error is most easily visible at 54 fs, where an additional Gaussian shaped wavepacket appears left of the nonlinearity. As the propagation continues and the wavepacket crosses the nonlinearity again, the error accumulates.

However, approaching the error generally from a theoretical perspective is difficult because $j$ is only defined for a coordinate transformation of equal numbers of coordinates. Since most grid based \ac{qd} approaches rely on reducing the number of reactive coordinates $\mathbf{q}$, the approximation of a constant $j$ gets easily conflated with the error introduced by removing all non-reactive coordinates not relevant for the studied system. There are two important steps to unravel this combined error related therewith. The first one is to clarify, under which conditions $j$ is actually constant. The second one is to quantify and reduce the error which is introduced by the removal of (non-reactive) coordinates but associated with the first step \cite{Alexandrov1998, Thallmair2016a}. 

A straightforward way to argue the first step is to take a look at the meaning of the Jacobian determinant $j$. It describes the volume change during a coordinate transformation. If we stay in the space of continuous well-behaved coordinate spaces that we need for our \ac{qd} calculations to work well, we can ensure this constancy for all cases where the dimensionality of a system is reduced. As the change in volume will also be continuous along the different coordinates and as long as there is a single non-reactive coordinate which we do not include in our subspace, we can simply define this additional coordinate as ensuring a constant $j$. One can think of the additional coordinate as stretching when the volume of the included reactive coordinates shrink and compressing when their volume grows.

Thus for all practical purposes $j$ can already be considered being constant under the assumption that the not included non-reactive coordinates take care of this by construction. However, it is not yet clear, which kind of error is introduced by the removal of those coordinates. In what follows, we will investigate this error further. Therefore we propose the following model scenario. If we construct a system whose dimensionality could be reduced without introducing any error, we can ensure that any error occurring due to a flawed reduction procedure is not an intrinsic effect of the dimensionality reduction itself. In other words, any error introduced to the correctly reduced system must also be present in an identical system that was created by reducing the dimensionality of the full system in a non-optimal way. This will allow us to isolate the error introduced by reducing different sets of coordinates.

The wavefunction $\Psi_0$ of our chosen full-dimensional system can be described by a product ansatz:
\begin{align}
 \Psi_0=\Psi_{\mathbf x}\Psi_{\mathbf y} \text{ .}
\label{eq3}
\end{align}
Here $\mathbf x$ denotes the part of coordinates that will be kept in the system of reduced dimensionality and $\mathbf y$ the part that will be removed. The Hamiltonian $H$ can be written as a sum of two separate Hamiltonians -- one for each coordinate subset $\mathbf x$ and $\mathbf y$:
\begin{align}
 H=H_\mathbf x+H_\mathbf y \text{ .}
\label{eq4}
\end{align}
That the product ansatz works can be seen by inserting Eqs. \ref{eq3} and \ref{eq4} into the Schr\"{o}dinger equation (Eq. \ref{eq:sg}) and separating the variables, leading to
\begin{align}
 \frac{1}{\Psi_\mathbf y}\left(i\frac{\partial}{\partial t} \Psi_\mathbf y-H_\mathbf y\Psi_\mathbf y\right)=C=-\frac{1}{\Psi_\mathbf x}\left(i\frac{\partial}{\partial t} \Psi_\mathbf x-H_\mathbf x\Psi_\mathbf x\right) \text{ .}
\label{eq5}
\end{align}
Since $C$ has to be a coordinate independent constant it is equivalent to some constant potential, which is irrelevant to the time evolution of the subsystems. From Eq. \ref{eq5} we can derive two separate Schr\"{o}dinger equations, propagating the subsystems $\mathbf x$ and $\mathbf y$ independently. Now, for all observables that only focus on results in the $\mathbf x$ subspace, we can write
\begin{align}
 \left<\Psi|O_\mathbf x|\Psi\right>= \left<\Psi_\mathbf x|O_\mathbf x|\Psi_\mathbf x\right> \left<\Psi_\mathbf y|\Psi_\mathbf y\right> \text{ .}
\label{eq6}
\end{align}
Since in Eq. \ref{eq6} all observables only depend on the $\mathbf x$ subsystem, the dimensionality reduction to only the coordinates $\mathbf x$ can be performed without introducing any error. As a result, we can now make the aforementioned comparison.

Case 1: We start from the reduced subsystem $\mathbf x$ and introduce a coordinate transformation to an reactive coordinate set $\mathbf s$. Since we do not reduce the dimensionality any further, any error introduced by $j$ not being constant is just the difference between Eq. \ref{eq1} and Eq. \ref{eq2}.

Case 2: We start from the full system and introduce a coordinate transformation between $\mathbf x$ and $\mathbf s$. This also results in a coordinate transformation between the full coordinate space in its original coordinates and the full chosen set of reactive and non-reactive coordinates, including those that ensure a constant $j$. Now the dimensionality is reduced by removing all degrees of freedom not included in $\mathbf s$.

In both cases the initial conditions for the propagation are identical and the time evolution after correctly performing the coordinate transformation will also be. We can therefore conclude that the easily quantifiable error in case 1 is indeed the error that will occur due to a non-constant $j$. This error could be eliminated by ensuring the transformation between $\mathbf x$ and $\mathbf s$ has a constant $j$.

\begin{figure} [ht!]
\begin{center}
   \includegraphics[width=0.6\columnwidth,keepaspectratio=true]{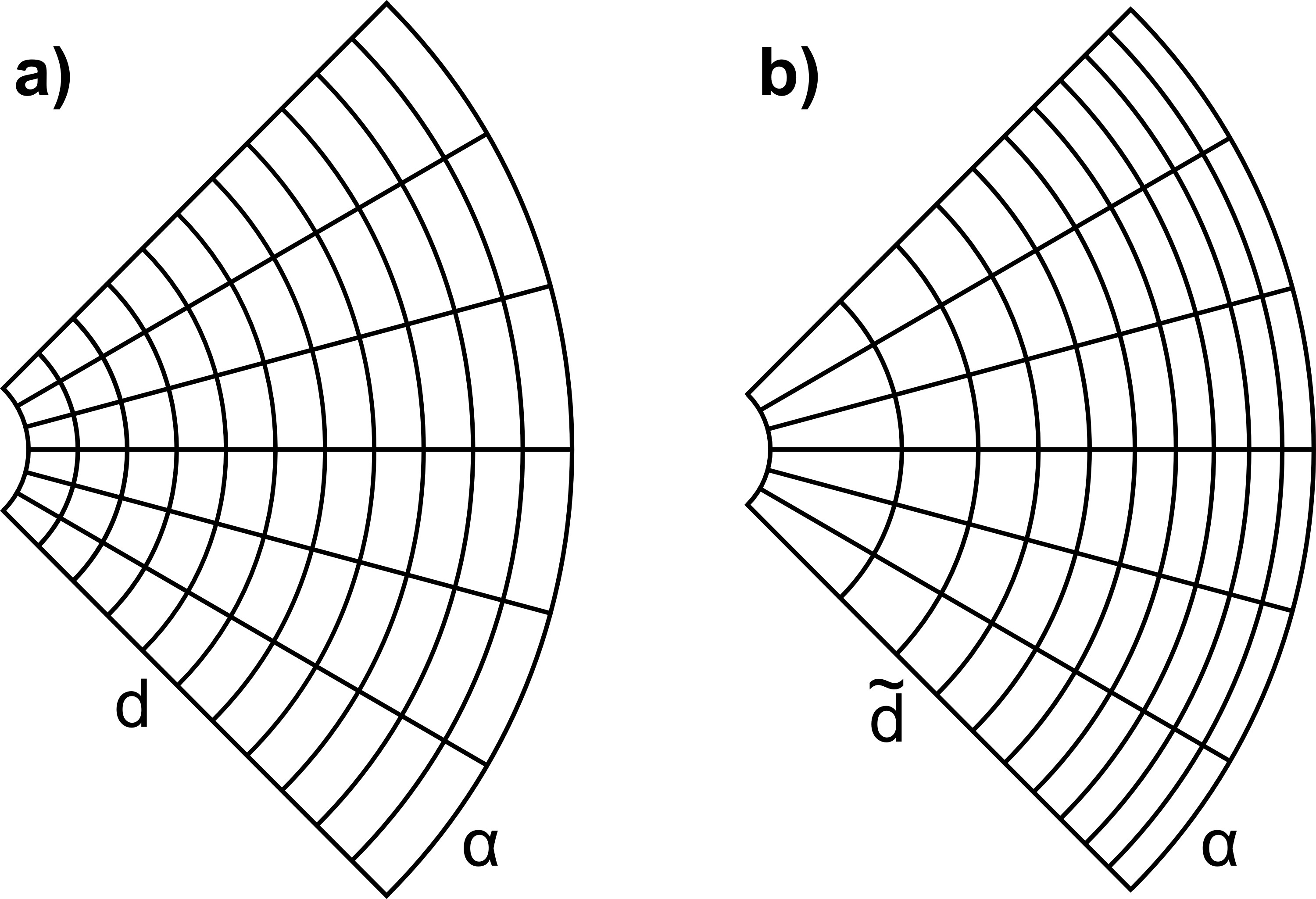} 
\end{center}
  \caption{Two types of coordinates sampling a distance and an angle. a) shows an equidistantly spaced grid using an angle $\alpha$ and a distance $d$ that does not have a constant $j$. b) shows an equidistantly spaced grid using the same angle $\alpha$ and another distance variable $\widetilde d$ transformed according to $\widetilde d=d^2$. In this case $j$ is constant.}
  \label{fig:grid}
\end{figure}

For molecular cases there are at least two view points. Since $j$ represents the change in volume due to the coordinate transformation, one way would be to rescale reactive coordinates so that the volume spanned by equispaced grid points remains constant at all positions. Let us consider an easy example with a bond distance $d$ and an angular coordinate $\alpha$. If $d$ and $\alpha$ are equidistantly spaced, the volume segments of the reactive coordinate space grid will increase linearly with $d$. In this case rescaling the distance coordinate $d$ to $\widetilde{d}=d^2$ will remove the errors by making $j$ constant. This example is visualized in Figure \ref{fig:grid}. It can be seen that in a) the volume elements spanned by the grid points differ in size, whereas they are all equal in b). In this example, it is obvious that the rescaling does not change the subspac. Thus, errors due to a non-constant $j$ are removed. A more rigorous treatment would be by ensuring that the Jacobian determinant $j$ between the reactive coordinates and its locally tangential Cartesian subspace remains constant. The benefits of this approach can be seen in cases where the tangential Cartesian subspace changes along the reactive coordinates. An example for this is a reactive subspace spanned by a bond angle and a corresponding dihedral angle. There is no single two-dimensional tangential space for this subspace and for every grid point the volume change under coordinate transformation has to be considered relative to its locally tangential Cartesian subspace. This volume change would then again needed to be kept constant using the appropriate rescaling procedures of the coordinates.

In summary, we showed that the approximation of a constant Jacobian determinant in the Wilson G-matrix formalism can lead to errors using non-linear coordinates in molecular \ac{qd}. However, the actual source of the error is hard to find as the coordinate reduction is commonly performed in the same step as the investigated approximation of a constant Jacobian determinant $j$. We presented a simple and straightforward solution to this problem: By ensuring a constant Jacobian determinant for the chosen subspace of the reactive coordinates $\mathbf q$ with respect to its tangential Cartesian subspace the approximation can be fulfilled and the only remaining error occurs due to the coordinate reduction itself.

In examples of molecular reactions, where a constant Jacobian determinant is not feasible -- e.g. due to limitations on the number of grid points to $\sim 10^9$ -- a solution might be to keep the Jacobian determinant between the non-linear reactive coordinate space and its locally tangential Cartesian subspace position dependent and to use it to substitute the full Jacobian determinant. We plan to investigate this hypothesis in more detail in the future.

\section*{Acknowledgement}
J.P.P.Z. and R.d.V.-R. thank the Deutsche Forschungsgemeinschaft for support via the excellence cluster Munich-Centre for Advanced Photonics (MAP) and the SFB749. S.T. thanks the European Commission for support via a Marie Sk\l{}odowska-Curie Individual Fellowship (project code 748895).

\bibliography{lit}

\end{document}